\documentclass[twocolumn]{aastex62}
\pdfoutput=1 
\usepackage{amsmath,amstext}
\usepackage{apjfonts} 
\usepackage[figure,figure*]{hypcap}
\usepackage{graphicx}	
\usepackage{amssymb}	
\usepackage{booktabs}
\usepackage{chemfig}

\received{}
\revised{}
\accepted{}
\submitjournal{ApJ}

\begin{document}

\title{H- and Dissociation in Ultra-hot Jupiters: A Retrieval Case Study of WASP-18b}

\correspondingauthor{Siddharth Gandhi}
\email{Siddharth.Gandhi@warwick.ac.uk}

\author{Siddharth Gandhi}
\affiliation{Institute of Astronomy, University of Cambridge, Madingley Road, Cambridge, CB3 0HA, UK}
\affiliation{Department of Physics, University of Warwick, Coventry CV4 7AL, UK}
\author{Nikku Madhusudhan}
\affiliation{Institute of Astronomy, University of Cambridge, Madingley Road, Cambridge, CB3 0HA, UK}
\author{Avi Mandell}
\affiliation{NASA Goddard Space Flight Center, Greenbelt, MA, USA}

\begin{abstract}
Atmospheres of a number of ultra-hot Jupiters (UHJs) with temperatures $\gtrsim$2000~K have been observed recently. Many of these planets show largely featureless thermal spectra in the near-infrared observed with the HST WFC3 spectrograph (1.1-1.7~$\mu$m) even though this spectral range contains strong H$_2$O opacity. Recent works have proposed the possibility of H- opacity masking the H$_2$O feature and/or thermal dissociation of H$_2$O causing its apparent depletion at the high temperatures of UHJs. In this work we test these hypotheses using observations of the exoplanet WASP-18b as a case study. We report detailed atmospheric retrievals of the planet using the HyDRA retrieval code, extended to include the effects of H- opacity and thermal dissociation. We report constraints on the H$_2$O, CO and H- abundances as well as the pressure-temperature profile of the dayside atmosphere for retrievals with and without H-/dissociation for each dataset. We find that the H$_2$O and H- abundances are relatively unconstrained given the featureless WFC3 spectra. We do not conclusively detect H- in the planet contrary to previous studies which used equilibrium models to infer its presence. The constraint on the CO abundance depends on the combination of WFC3 and Spitzer data, ranging from solar to super-solar CO values. We additionally see signs of a thermal inversion from two of the datasets. Our study demonstrates the potential of atmospheric retrievals of UHJs including the effects of H- and thermal dissociation of molecules. 
\end{abstract}

\section{Introduction}\label{sec:intro}

Recent observations of hot Jupiters are providing some of the most precise atmospheric spectra of exoplanets to date. Such data have provided important constraints on the atmospheric parameters, e.g. composition and temperature profile, thanks to instruments such as the HST WFC3 spectrograph \citep{deming2013, madhu2019}. A new class of irradiated giant planets, that of ultra-hot Jupiters (UHJs), has emerged recently. These planets have dayside temperatures exceeding $\sim$ 2000~K owing to strong incident stellar flux due to their short orbital periods. Observations of such objects have revealed that many of these atmospheres show no significant features in the WFC3 bandpass \citep[e.g.][]{sheppard2017, kreidberg2018, arcangeli2018, lothringer2018, mansfield2018, arcangeli2019}. With such featureless spectra, constraints on species such as H$_2$O have been minimal, and only upper limits have generally been derived. 

Several hypotheses have been put forward to explain the lack of strong H$_2$O features and generally featureless spectra for UHJs. A high C/O ratio has been invoked as a possible explanation for muted H$_2$O features in many hot Jupiters, both in transmission and emission \citep[e.g.][]{madhu2011, madhu2014, sheppard2017, pinhas2019}. Previous studies have shown that a super-solar atmospheric C/O ratio can deplete H$_2$O in the atmosphere for temperatures $\gtrsim 1200$~K \citep[e.g.,][]{madhu2012, moses2013, drummond2019}. In this case the oxygen preferentially binds to form CO when C/O $\gtrsim$1, thus significantly reducing the overall H$_2$O abundance. The low H$_2$O abundance, in turn, can lead to weaker features in the HST WFC3 band (1.1-1.7~$\mu$m) where H$_2$O has strong opacity.

An alternate explanation is the thermal dissociation of H$_2$O in UHJs. Equilibrium models of UHJs have revealed that the dissociation of species such as H$_2$O and the formation of H- may play an important role in determining the emergent spectrum \citep{arcangeli2018, parmentier2018, lothringer2018}. The H- is formed in irradiated environments rich in hydrogen and free electrons, and is in fact the dominant source of opacity in stars cooler than 7000~K \citep{wishart1979, bell1987}. The H- ion possesses broad continuum opacity in the 1.1-1.7~$\mu$m WFC3 band. This opacity can fill in the gap in between the H$_2$O feature at $\sim$1.4~$\mu$m and result in featureless spectra \citep{parmentier2018}. In addition, at high temperatures thermal dissociation of H$_2$O would act to diminish the H$_2$O abundance and further mute H$_2$O features below expected levels \citep{arcangeli2018}. Very high temperatures are required for both H- opacity and thermal dissociation to be significant, but UHJs such as WASP-121b, WASP-18b, WASP-12b, Kepler-13Ab, HAT-P-7b and KELT-9b have equilibrium temperatures in excess of $\sim$2000~K and therefore warrant consideration of these contributions \citep[e.g.][]{arcangeli2018, kreidberg2018, mansfield2018, lothringer2018, kitzmann2018}. Thermal dissociation is generally also favoured by lower pressures \citep[e.g.][]{parmentier2018}. However, it is unclear whether the absence of an H$_2$O feature in the WFC3 bandpass for UHJs can be solely attributed to these two factors - some UHJs with similar day-side temperatures (e.g. KELT-1b, Kepler-13Ab) do in fact show strong H$_2$O absorption \citep{beatty2017, beatty2017_kelt1, parmentier2018}.

In this work we assume an agnostic position \emph{a priori} on which effect is dominant in UHJs. We investigate evidence for the above hypotheses by performing spectral retrievals of a well-known UHJ, WASP-18b \citep{sheppard2017,arcangeli2018}. Previous retrievals of WASP- 18b have proposed a high C/O ratio as an explanation for the observed spectrum, which does not show strong H$_2$O features in the WFC3 band \citep{sheppard2017}. However, their derived CO abundance implied a very high metallicity of 283$^{+395}_{-138}$ $\times$ solar. On the other hand, \citet{arcangeli2018} explored a grid of equilibrium models to propose H- and dissociation as a potential explanation for the featureless spectrum. General circulation models have also been used to explain the phase curve of WASP-18b \citep{arcangeli2019}. We assess evidence for the two proposed hypotheses by conducting retrievals on the three HST WFC3 datasets available for this planet \citep{sheppard2017, arcangeli2018, arcangeli2019} along with the associated Spitzer photometry between 3.6-8~$\mu$m \citep{nymeyer2011, maxted2013, sheppard2017}.

We explore different proposed scenarios to explain the emission spectrum of WASP-18b. We retrieve the abundances of species such as H$_2$O and CO as free parameters rather than enforcing chemical equilibrium which other studies have done. Our parametric temperature profile explores isothermal as well as inverted and non-inverted profiles to determine the best fit to the observations. We additionally include thermal dissociation and H- opacity in our model to explore their effects on the retrieved parameters both separately and in tandem over multiple separate retrievals. We perform extensive retrievals for three HST WFC3 datasets for the dayside emission spectrum of WASP-18b, namely from \citet{sheppard2017}, \citet{arcangeli2018} and \citet{arcangeli2019}. All of these datasets represent a measurement of the dayside flux from WASP-18b, but \citet{arcangeli2019} produce their spectral dataset by binning the phase-resolved spectra between phases of 0.4-0.45 and 0.55-0.6, just before and after secondary eclipse, rather than determining the best estimate of the spectrum during eclipse itself, as is standard in secondary eclipse analyses. Therefore the \citet{arcangeli2019} dataset is probing a slightly different portion of the dayside disk of the planet compared with the other two datasets. The bulk planetary conditions may be roughly similar between the datasets, though an exact comparison of the retrieved properties from the \citet{arcangeli2019} dataset and the other two datasets is not our goal.

In what follows, we describe the modelling and retrieval method in section \ref{sec:methods}. This is followed by the results for the retrievals conducted with and without H- and dissociation. We discuss the retrieved parameters and compare and contrast each retrieval. Finally we discuss the conclusions and future directions in section \ref{sec:wasp18_conclusion}. 

\section{Methods}\label{sec:methods}

Here we describe the methods used for the retrieval of the emission spectra of UHJs. We build upon the HyDRA retrieval code \citep{gandhi2018} and generalise it to UHJs. In particular, the HyDRA framework is extended to model the atmospheres of UHJs by the inclusion of thermal dissociation and H- opacity. Previous studies have shown the importance of these at temperatures in excess of $\sim$2000~K \citep{arcangeli2018, parmentier2018, lothringer2018}. The additions to the retrieval model are discussed below in sections \ref{sec:dissoc_method} and \ref{sec:h-opac}.

The composition and pressure-temperature (P-T) profile are free parameters in the model. For the chemical composition we do not make any assumptions of chemical equilibrium, instead derive the abundances with uniform priors with no a priori constraints. The pressure-temperature (P-T) profile has been parameterised using the method outlined in \cite{madhu2009}. The 6 free parameters which parameterise the P-T profile freely allow for inversions in the atmosphere. Through these parameters we are able to explore a wide range of inverted, non-inverted and isothermal profiles to best fit the observations. We also assume a plane-parallel geometry with the assumption that the atmosphere is in hydrostatic equilibrium and local thermodynamic equilibrium. The emergent flux out of the atmosphere is calculated from double-ray integration of emergent rays of radiation out of the atmosphere as discussed in \citet{gandhi2018}. To generate the stellar spectrum we use the Kurucz model grid \citep{Kurucz_1979_paper}. The Bayesian parameter estimation is performed using the Nested Sampling algorithm implemented in the MultiNest package \citep{feroz2008, feroz2009, buchner2014}. Further details on the retrieval and the model setup can be found in \citet{gandhi2018}.

The chemical opacities have been derived using high temperature line lists for the various species. These species include H$_2$O, CO, CO$_2$, TiO and VO, which are expected to be prominent in ultra-hot H$_2$ dominated atmospheres \citep{madhu2012}. The line lists for H$_2$O, CO and CO$_2$ have been obtained from the HITEMP database \citep{rothman2010}. The high temperature line lists for TiO and VO have been obtained from the ExoMol database \citep{tennyson2016, mckemmish2016, mckemmish2019}. The opacity due to the presence of H- has been calculated following \cite{john1988} and \cite{wishart1979} (see section \ref{sec:h-opac}). Each molecular line has been spectrally broadened by temperature and pressure resulting in a Voigt profile as a function of wavelength \citep{gandhi2017}. The molecular cross sections for each of the prominent species has been precomputed at a wavenumber spacing of 0.1~cm$^{-1}$ between 0.4-50~$\mu$m with temperatures and pressures ranging from 300-3500~K and 10$^2$-10$^{-5}$~bar respectively. We additionally include opacity from collisionally induced absorption (CIA) from H$_2$-H$_2$ and H$_2$-He interactions \citep{richard2012}.

\begin{figure}[ht]
    \centering
	\includegraphics[width=\columnwidth,trim=0.2cm 0 0.2cm 0,clip]{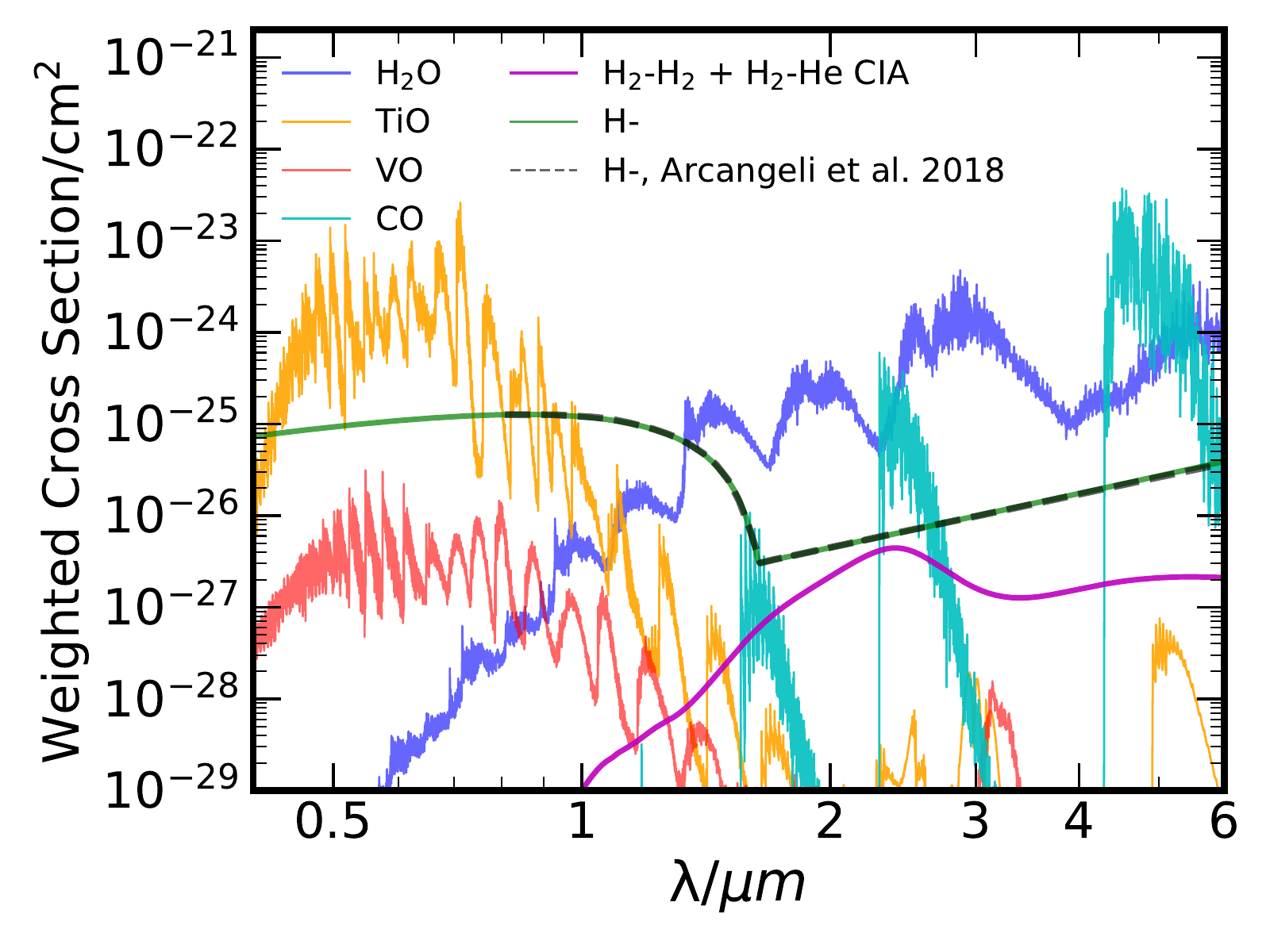}
    \caption{Abundance weighted cross sections for the prominent species in our retrieval at a temperature of 2900K and 0.33~bar pressure. The abundances have been calculated from the dissociation model by \citet{parmentier2018} and the cross sections derived in \cite{gandhi2017}. The H- opacity is calculated in section \ref{sec:h-opac}. We also show the abundance weighted H- cross section from \citet{arcangeli2018}.}
    \label{fig:wasp18cs}
\end{figure}

\subsection{Thermal Dissociation}\label{sec:dissoc_method}

Ultra-hot giant planets have temperatures in excess of 2000~K, where many molecular species may begin to thermally dissociate \citep{arcangeli2018, lothringer2018, parmentier2018}. \citet{arcangeli2018} in particular have argued that temperatures in excess of $\sim$2000-2500~K can result in the thermal dissociation of molecular species. This may result in significant depletion of species such as H$_2$O in the upper atmosphere, which acts to lower the infrared photosphere deeper into the atmosphere. In addition, the depletion in the upper atmosphere can also reduce the extent of the $\tau_\nu \sim 1$ surface \citep{parmentier2018}. This can thus reduce the extent of spectral features and therefore significantly alter the emission spectra for such ultra-hot planets. This has been used to explain many of the featureless WFC3 spectra for some of the hottest systems such as WASP-121b, WASP-18b and WASP-103b \citep{parmentier2018, arcangeli2018, kreidberg2018}. We therefore include thermal dissociation for H$_2$O, TiO, VO and H-, which are expected to be susceptible to dissociation at such high temperatures.

We calculate the volume mixing fractions of these species as a function of pressure and temperature from the dissociation model developed by \citet{parmentier2018}. The volume mixing fraction $A_i(P,T)$ of species $i$ is given by
\begin{align}
    \frac{1}{A_i^{0.5}} &= \frac{1}{A_{0,i}^{0.5}} + \frac{1}{A_{d,i}^{0.5}}.\label{eqn:dissoc}
\end{align}
$A_{0,i}$ is the deep atmosphere (undissociated) abundance and 
\begin{align}
    \log(A_{d,i}) &= \alpha_i \log(P) + \beta_i/T -\gamma_i,\label{eqn:Ad}
\end{align}
for pressure P(bar) and temperature T(K) and parameters $\alpha_i$, $\beta_i$ and $\gamma_i$, which are fits to the equilibrium abundances at solar composition \citep{parmentier2018}. Using the deep atmosphere abundance $A_{0,i}$ for each species $i$ we calculate the dissociated volume mixing ratio $A_i(P,T)$ according to the pressure and temperature through equations \ref{eqn:dissoc} and \ref{eqn:Ad}. This is done for each layer of the model atmosphere resulting in a volume mixing fraction that is a function of atmospheric depth. Thermal dissociation of CO has not been included here as it does not readily dissociate at the temperatures seen on WASP-18b and other ultra-hot Jupiters \citep[e.g.][]{lodders2002, arcangeli2018, parmentier2018}. In our retrieval the undissociated abundance priors for H$_2$O, TiO, VO and H- span $\log(\mathrm{A_{0,i}}) = -15$ and $0$. Our upper bound for the H- prior is restricted to $\log(\mathrm{H-}) = -7$ in our fiducial retrieval given that equilibrium models assuming solar composition predict a lower abundance of $\log(\mathrm{H-}) \approx -8.3$ \citep[e.g.][]{parmentier2018}. This parametric model for thermal dissociation is most accurate near solar composition but we should highlight the fact that our retrieval explores a wider range of composition. The dissociated H- abundance in particular may be inaccurate away from solar abundance chemistry as it depends strongly on the free electron abundance \citep{arcangeli2018, parmentier2018}.
In addition, we have also run retrievals including the effect of H$_2$ dissociation to study its effect on the retrieved parameters.

\subsection{H- Opacity}\label{sec:h-opac}

At the high temperatures of UHJs, H- may be present in the atmosphere and thus affect the emergent spectrum due to its strong cross section in the near infrared \citep{arcangeli2018, parmentier2018, lothringer2018}. The H- cross section arises as a result of two separate sources of opacity, bound-free photo-detachment and free-free transitions. Both of these sources of opacity result in broad features in the cross section, unlike with molecular cross sections which consist of many millions of broadened transition lines. The bound-free transitions have strong absorption at $\lambda$<1.64~$\mu$m, the photo-detachment threshold wavelength, and are in fact the dominant source of opacity in cool stars \citep{wishart1979}. At longer wavelengths, the free-free transitions are the only contributors to the cross section of H-. Both of these opacity contributions are summed to give the total cross section of H-. Figure~\ref{fig:wasp18cs} shows the abundance weighted cross section at a representative temperature of 2900~K and 0.33~bar pressure. The abundances are calculated from the dissociation model in section \ref{sec:dissoc_method} assuming a solar composition atmosphere.

The bound-free opacity arises due to the absorption of a photon by H-,
\begin{equation}
\begin{split}
h\nu + \mbox{H-} &\rightarrow \mbox{H} + \mbox{e}^-.
\end{split}\nonumber
\end{equation}
The cross section of this reaction, $\sigma_\mathrm{H-}$, is given by
\begin{align}
    \sigma_{\rm H-}(\lambda < \lambda_0) &= 10^{-18} \lambda^3 \bigg(\frac{1}{\lambda} - \frac{1}{\lambda_0}\bigg)^{3/2} f(\lambda) {\rm cm^2},
\end{align}
where $\lambda$ represents the wavelength in $\mu$m, $\lambda_0 = 1.6419 \, \mu$m is the photo-detachment threshold and $f(\lambda)$ is a slowly varying function \citep{john1988}. These bound-free transitions result in a cutoff in the opacity at the photo-detachment threshold, with a strong cross section at $\lambda<\lambda_0$. This cutoff is of particular importance given that $\lambda_0$ is within the HST WFC3 bandpass (1.1-1.7~$\mu$m). 

The free-free transitions on the other hand occur due to the following chemical interaction \citep{bell1987}
\begin{equation}
\begin{split}
h\nu + \mbox{e}^- + \mbox{H} &\rightarrow \mbox{H} + \mbox{e}^-.
\end{split}\nonumber
\end{equation}
These transitions are important sources of infrared opacity for late spectral-type stars \citep{bell1987}. This opacity may also be important in UHJs, particularly for the Spitzer photometric observations at $\lambda > $3~$\mu$m given that this cross section increases with wavelength. A more detailed study of the free-free absorption due to H- can be found in \cite{john1988} and \cite{bell1987}. In our model the combined opacity for both bound-free and free-free transitions is included, and we leave the volume mixing ratio of H- as a free parameter in the retrieval which is used to compute the total opacity due to H-, including the effect of thermal dissociation as described earlier.

\begin{table}[ht]
    \centering
    \begin{tabular}{c|c}
        \textbf{WASP-18 System Parameter} & \textbf{Assumed Value in Retrieval} \\
        \hline
        Planetary Radius / R$_\mathrm{J}$ & 1.19\\
        Planetary Mass / M$_\mathrm{J}$ & 10.4 \\
        Stellar Effective Temperature / K & 6430.0 \\
        Stellar Radius / R$_\mathrm{\odot}$ & 1.26\\
        Stellar Mass / M$_\mathrm{\odot}$ & 1.46\\
        Stellar Metallicity $[\mathrm{Fe/H}]$ & 0.10 \\
    \end{tabular}
    \caption{The values for the WASP-18 system used in this work. We assume that the radius of Jupiter is 7.15$\times 10^7$m.}
    \label{tab:wasp18_params}
\end{table}

\begin{table}[ht]
    \centering
    \begin{tabular}{c|c}
        \textbf{Parameter} & \textbf{Prior Range} \\
        \hline
        $\log(\mathrm{H}_2\mathrm{O})$, $\log(\mathrm{CO})$, $\log(\mathrm{CO}_2)$, $\log(\mathrm{TiO})$, $\log(\mathrm{VO})$ & -15 - 0 \\
        $\log(\mathrm{H-})$ & -15 - -7$^*$ \\
        $T_{100\mathrm{mb}}$/K & 300 - 4000 \\
        $\alpha_1$/$\mathrm{K}^{-1/2}$ & 0 - 1 \\
        $\alpha_2$/$\mathrm{K}^{-1/2}$ & 0 - 1 \\
        $\log(P_1$/bar) & -6 - 2\\
        $\log(P_2$/bar) & -6 - 2\\
        $\log(P_3$/bar) & -2 - 2\\
    \end{tabular}
    \caption{The priors for the parameters in the retrievals of WASP-18b carried out in this work. Each of these priors are uniform in the range given. \newline $^*$ We also run retrievals extending the undissociated H- prior to $\log(\mathrm{H-})$ = 0.}
    \label{tab:ret_params}
\end{table}

\subsection{Retrieval Setup}

\begin{figure*}[ht]
	\includegraphics[width=\textwidth,trim=1.7cm 0 2cm 0,clip]{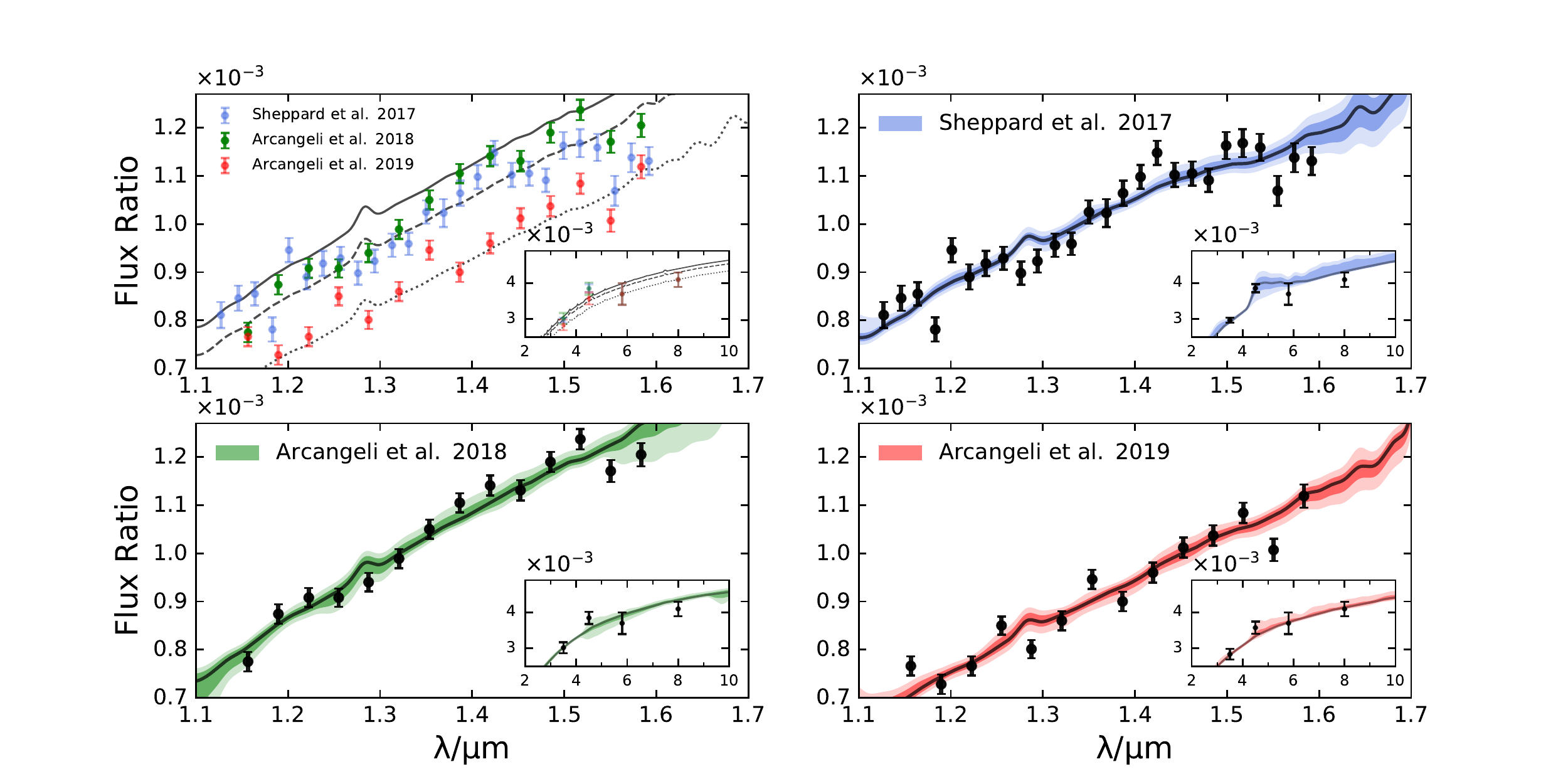}
    \caption{Thermal emission spectra (planet/star flux ratios) of WASP-18b from the different datasets and model fits from retrievals. The main panel shows the HST WFC3 data (1.1-1.7~$\mu$m) and the inset in each panel shows the Spitzer IRAC photometric points. The top left panel shows the three datasets together with black body curves at 2800~K (dotted line), 2900~K (dashed line) and 2950~K (solid line). The other three panels show the observed spectra from each dataset and corresponding model fits from retrievals with H- and dissociation. The dark and light colours indicate the 1$\sigma$ and 2$\sigma$ uncertainty respectively, and the solid line shows the median best fit curve for each. The final three HST WFC3 points for the \citet{sheppard2017} dataset have not been included into our retrievals for comparisons given that the other two datasets do not have data at these wavelengths.}
    \label{fig:wasp18_flux}
\end{figure*}

\begin{figure*}[ht]
	\includegraphics[width=\textwidth,trim=6.3cm 0 4.6cm 0,clip]{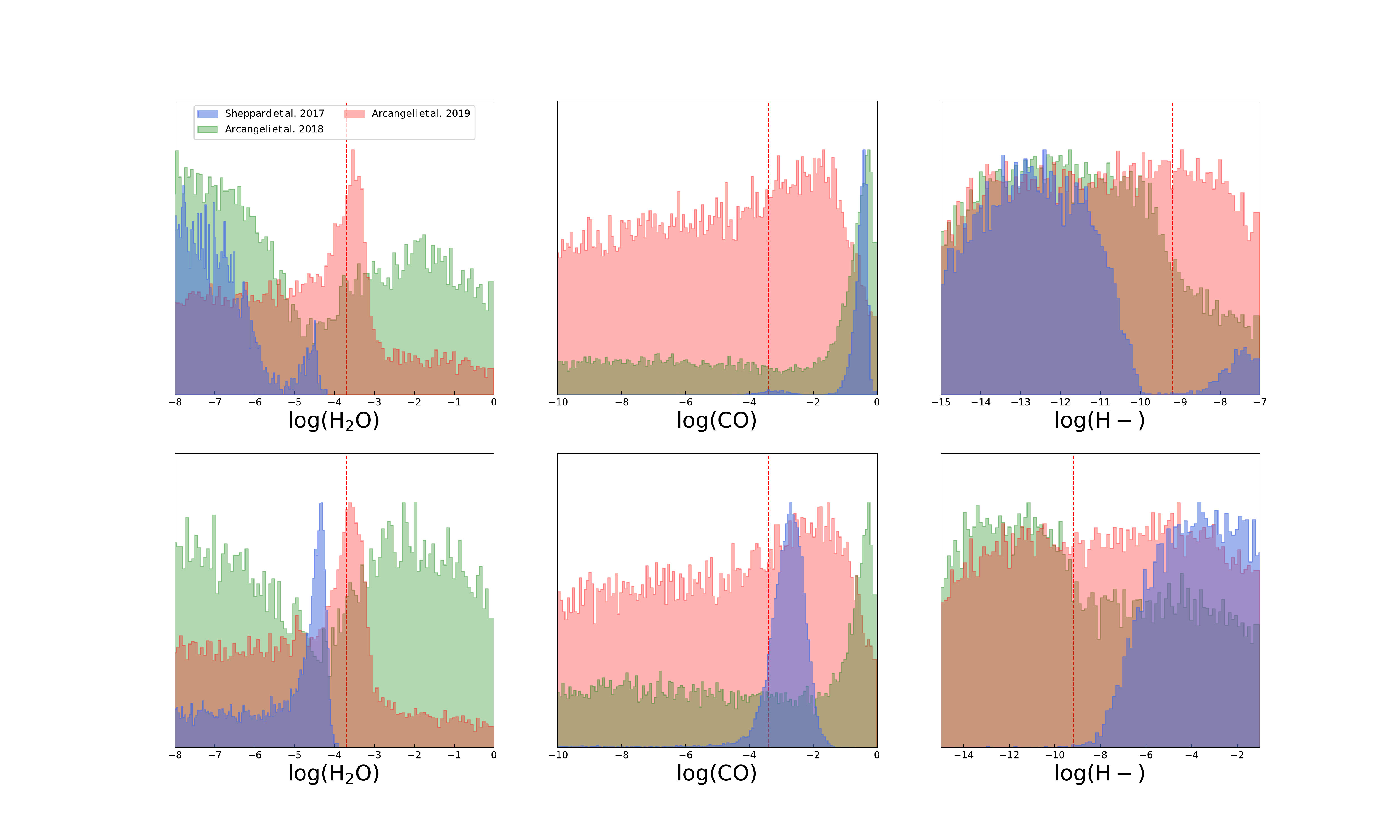}
    \caption{Posterior probability distributions of undissociated H$_2$O, CO and H- abundances for the various retrievals conducted. The top panels show the fiducial case where the H- prior was restricted to be between $\log(\mathrm{H-}) = -15$ and $-7$, whereas the bottom panel shows the wide prior case where the upper H- prior was extended to $\log(\mathrm{H-}) = -0$. In each retrieval shown the H$_2$O and H- were thermally dissociated with pressure and temperature according to the model in section \ref{sec:dissoc_method}. The red dashed line shows the expected abundance for each of the species assuming chemical equilibrium with solar elemental abundances \citep{parmentier2018}.}
    \label{fig:wasp18histogram}
\end{figure*}

We analyse both HST WFC3 and Spitzer observations (across the 4 channels) for WASP-18b using all of the three available dayside observations \citep{sheppard2017, arcangeli2018, arcangeli2019}. These HST and Spitzer observations are some of the most precise measurements of ultra-hot Jupiter emission available today. We note, as discussed in section~\ref{sec:intro}, that the \citet{arcangeli2019} dataset is derived from a slightly different part of the dayside disk of the planet and hence the retrieved properties from this spectrum may not be directly comparable to those from the other two datasets. Our retrievals for the \citet{sheppard2017} dataset do not include the final three data points in the WFC3 range, as \citet{arcangeli2018} and \citet{arcangeli2019} do not include data beyond 1.6~$\mu$m. This allows us to consistently retrieve each dataset over the same range of observations. The \citet{arcangeli2019} dataset has been reconstructed adopting conservative error bars based on the \citet{arcangeli2018} data. The 3.6~$\mu$m and 4.5~$\mu$m Spitzer data for the \citet{arcangeli2018} and \citet{arcangeli2019} datasets are obtained from \citet{maxted2013}, but \citet{sheppard2017} re-analysed these photometric data for their work and achieved significantly smaller uncertainties. For all three studies the eclipse depths in the 5.8~$\mu$m and 8~$\mu$m channels were obtained from previous analyses by \citet{nymeyer2011}.

For each retrieval we use 4000 evenly spaced wavelength points between 1.1-10.4~$\mu$m for our line-by-line calculations, encompassing both the WFC3 and Spitzer IRAC observations. The assumed planetary and stellar parameters are given in table \ref{tab:wasp18_params} and obtained from \citet{shporer2018}. To generate the stellar spectrum we use the Kurucz model grid \citep{Kurucz_1979_paper, kurucz_model}. Each retrieval contains 6 free parameters for the P-T profile and 5 for the volume mixing fraction of H$_2$O, CO, CO$_2$, TiO and VO. The priors for these parameters are given in table \ref{tab:ret_params}. An additional free parameter for the H- abundance is also included in some of the retrievals. H- is expected to be present at $\log(\mathrm{H-}) \approx -8.3$ in the dayside atmospheres of UHJs \citep[e.g.][]{parmentier2018, arcangeli2018}. Our fiducial case assumes an undissociated H- prior range $\log(\mathrm{H-}) = -15$ to $-7$, but we have also run retrievals with a wide H- prior extending this to $\log(\mathrm{H-}) = 0$. The retrieval without H- or dissociation is similar to previous work \citep{sheppard2017} but we do not include CH$_4$ or HCN given that no evidence was seen for either of these species. In order to accurately retrieve the parameters and provide Bayesian evidences we use the Nested Sampling algorithm MultiNest \citep{feroz2008, feroz2009, buchner2014}. Further details on the forward model calculations and retrieval framework can be found at the beginning of section \ref{sec:methods} and in \cite{gandhi2018}. 
\newline
\section{Results}\label{sec:wasp18_results}

\begin{table*}[ht]
    \centering
    \renewcommand{\arraystretch}{1.5}
    \begin{tabular}{l|l|cccc|l}
    \textbf{Dataset} &\textbf{Retrieval} & \textbf{$\rm{\mathbf{T_{100mb}}}$/K} & \textbf{$\rm{\mathbf{\log({H_2O})}}$} & \textbf{$\rm{\mathbf{\log(CO)}}$} & \textbf{$\rm{\mathbf{\log(H-)}}$}& \textbf{Model Comparison} \\
    \hline
    \citet{sheppard2017} & No H-, No dissoc. & ${2829}^{+36}_{-48}$ & <${-5.9}$ & ${-0.58}^{+0.24}_{-0.35}$ & - & Reference value\\
    
    & Dissoc., no H- & ${2805}^{+32}_{-44}$ & <${-6.1}$ & ${-0.41}^{+0.14}_{-0.25}$ & - & Disfavoured by 2.0$\sigma$\\
    
    & H-, No dissoc. & ${2804}^{+47}_{-99}$ & <${-0.70}$ &
    ${-1.10}^{+0.60}_{-6.8}$ & <${-3.9}$ & Disfavoured by 2.6$\sigma$\\
    
    & H-, dissoc. (wide prior) & ${3005}^{+31}_{-23}$ & ${-7.3}^{+2.8}_{-4.0}$ & ${-2.70}^{+0.45}_{-0.57}$ &  ${-3.4}^{+2.1}_{-2.2}$ & Disfavoured by 3.1$\sigma$\\
    
    & H-, dissoc. (fiducial) & ${2810}^{+33}_{-37}$ & <${-5.7}$ & ${-0.41}^{+0.15}_{-0.28}$ & <${-7.6}$ & Disfavoured by 2.4$\sigma$\\
    \hline
    \citet{arcangeli2018} & No H-, No dissoc. & ${2849}^{+64}_{-46}$ & <${-0.7}$ & ${-1.2}^{+1.0}_{-7.0}$ & - & Reference value\\
    
    & Dissoc., no H- & ${2858}^{+59}_{-55}$ & <${-0.5}$ & ${-2.5}^{+2.2}_{-6.8}$ & - & Disfavoured by 1.6$\sigma$\\
    
    & H-, No dissoc. & ${2868}^{+53}_{-54}$ & <${-0.6}$ & ${-2.0}^{+1.8}_{-6.4}$ &  <${-3.1}$ & Disfavoured by 1.6$\sigma$\\
    
    & H-, dissoc. (wide prior) & ${2907}^{+13}_{-67}$ & <${-0.4}$ & ${-5.4}^{+4.7}_{-4.7}$ &  <${-1.5}$ & Disfavoured by 1.9$\sigma$\\
    
    & H-, dissoc. (fiducial) & ${2889}^{+30}_{-78}$ & <${-0.5}$ & ${-4.0}^{+3.6}_{-5.6}$ &  <${-7.5}$ & Disfavoured by 1.5$\sigma$\\
    \hline
    \citet{arcangeli2019} & No H-, No dissoc. & ${2803}^{+18}_{-31}$ & ${-5.6}^{+3.3}_{-4.4}$ & $<{-0.3}$ & - & Reference value\\
    
    & Dissoc., no H- & ${2819}^{+12}_{-11}$ & ${-6.5}^{+3.1}_{-4.0}$ & $<{-0.3}$ & - & Disfavoured by 1.3$\sigma$\\
    
    & H-, No dissoc. & ${2800}^{+20}_{-33}$ & ${-4.9}^{+3.0}_{-4.5}$ & $<{-0.3}$ &  <${-2.5}$ & Favoured by 1.1$\sigma$\\
    
    & H-, dissoc. (wide prior) & ${2819}^{+12}_{-10}$ & ${-6.5}^{+3.0}_{-4.1}$ & $<{-0.3}$ & <${-1.4}$ & Disfavoured by 1.3$\sigma$\\
    
    & H-, dissoc. (fiducial) & ${2819}^{+12}_{-10}$ & ${-6.5}^{+3.1}_{-4.1}$ & $<{-0.3}$ &  <${-7.3}$ & Disfavoured by 1.3$\sigma$\\

\end{tabular}
    \caption{Retrieved temperature and abundance parameters for the various retrievals conducted on the emission spectrum of WASP-18b. In each case where there was a peak in the posterior distribution the retrieved median value is shown with its 1$\sigma$ uncertainty and where there was no observable peak the 2$\sigma$ upper limit is shown. We report median values for the abundances as conservative estimates, even though the modal values are significantly higher than the median values in some cases. In models with thermal dissociation the 100~mbar abundance is shown. The H- prior in the fiducial case was restricted to be between $\log(\mathrm{H-}) = -15$ and $-7$, whereas in the wide prior and H- without dissociation case the H- range was between $\log(\mathrm{H-}) = -15$ to $-0$. We also show the Bayesian comparison for each model versus our reference case.
    }
    \label{tab:retrievals}
\end{table*}

Here we discuss the results from retrievals of the emission spectrum of WASP-18b as considered by \citet{sheppard2017}, \citet{arcangeli2018} and \citet{arcangeli2019}. We retrieved the WFC3 and Spitzer data with and without H- and dissociation included in the retrieval model to see the effect of each separately and together for each of the available datasets. For each dataset we perform 5 retrievals in the various combinations (see Table \ref{tab:retrievals}). Our fiducial model includes dissociation and H- with the undissociated prior for H- restricted to be from $\log(\mathrm{H-}) = -15$ to $-7$. The posterior distributions for these retrievals are given in the appendix. Our unrestricted wide prior retrieval with H- and thermal dissociation considers an undissociated H- prior of $\log(\mathrm{H-}) = -15$ to $0$. Table \ref{tab:retrievals} summarises the constraints on the various atmospheric parameters from these retrievals, showing the P=100$\mathrm{mb}$ temperature, H$_2$O, CO and H- abundances for each dataset. The spectral fit for the fiducial retrieval is shown for each dataset in Figure~\ref{fig:wasp18_flux}. 

Figure~\ref{fig:wasp18_flux} shows a well-fit planet/star flux ratio for both the WFC3 and Spitzer observations (inset in the figure). The top left panel shows the datasets, along with black body curves for WASP-18b for various temperatures. The HST WFC3 points for all three show a relatively featureless spectrum, and thus the constraints on the spectroscopically active species in this spectral range are minimal as shown in Figure~\ref{fig:wasp18histogram}. The H- abundance shows a bimodal distribution for the \citet{sheppard2017} dataset. This is because the dip in the spectrum at the red end of the WFC3 observations, as shown in Figure~\ref{fig:best_fit}, can be explained by strong absorption due to either CO or H-. The Spitzer data for the \citet{sheppard2017} and \citet{arcangeli2018} datasets additionally constrain CO due to its feature in the 4.5~$\mu$m band. The CO$_2$, TiO and VO abundances do not show any significant constraints from the current observations given their likely trace abundances and weak features at the wavelengths probed.

The P-T profiles for each of the retrievals also agree well with each other, with the most well constrained temperature for each of the datasets in the photosphere at P$\sim$100~mbar as shown in Figure~\ref{fig:wasp18_PT}. This is as we would expect given the similarity of the data. For each dataset we find T$_\mathrm{100mb}$ between $\sim$2800-3000~K as expected from the black body curves in the top left panel of Figure~\ref{fig:wasp18_flux}. These also agree remarkably well with the expected equilibrium temperature of WASP-18b without significant redistribution (T$\approx$2850~K). We see a slight thermal inversion with the \citet{sheppard2017} and \citet{arcangeli2019} datasets, but a more isothermal profile with the \citet{arcangeli2018} data. However, the Arcangeli et al. datasets do allow for inverted and non-inverted temperature profiles due to uncertainties in the retrieved P-T profile. We will now discuss some key differences between the retrievals along with the implications below.

\subsection{H$_2$O Abundances}

We only retrieve weak constraints and upper limits for the H$_2$O abundance, even with the inclusion of H- and dissociation. This is due to the relatively featureless spectrum seen in all three datasets. Figure~\ref{fig:wasp18_flux} shows the fit to the 1.1-1.7~$\mu$m WFC3 data. At these wavelengths the opacity from H$_2$O, along with perhaps H-, is expected to be dominant over any other. Figure~\ref{fig:wasp18histogram} shows the retrieved H$_2$O abundance (undissociated) at a reference pressure of $P=100$~mbar. All of the retrievals that were run across all of the datasets do not show strong constraints on H$_2$O due to the featureless blackbody-like WFC3 spectra (see Table \ref{tab:retrievals}). Despite some tentative peaks in the posterior distributions H$_2$O cannot be conclusively detected in our retrievals. The inclusion of H- or dissociation into our model does not alter the H$_2$O abundances and all still have wide uncertainties on the abundance. In fact, as a general trend across all of the datasets, the inclusion of H- and dissociation increases the uncertainty on the retrieved abundances slightly due to the extra parameter.

The \citet{sheppard2017} data show upper limits on the H$_2$O at sub-solar abundance, but there is no definitive detection of H$_2$O. This is in agreement with retrievals published in their original study. Our new retrievals with H- and dissociation also only show a sub-solar upper limit on the H$_2$O. The fiducial case in Figure~\ref{fig:wasp18histogram} does show a peak at $\log(\mathrm{H_2O}) \sim -4.5$, but with a long tail of uncertainty extending to very low mixing ratios. However, all except one of the retrievals show no probability density for H$_2$O at solar or super-solar abundance for the \citet{sheppard2017} dataset. Such a sub-solar abundance, even with the inclusion of thermal dissociation, may have important consequences for the planetary atmosphere (see section \ref{sec:wasp18_conclusion}). The one exception is the case with H- and without dissociation, which indicates no significant peak at any abundance and constrains the 2$\sigma$ upper limit for H$_2$O to be $\log(\mathrm{H_2O}) \sim -0.7$ (see Table \ref{tab:retrievals}).

Retrievals with \citet{arcangeli2018} do not show any constraints on H$_2$O for any of the five cases run (see Table \ref{tab:retrievals}). The H$_2$O abundance retrieved from the H- and dissociation models is shown in Figure~\ref{fig:wasp18histogram} and indicates no significant peak at any abundance. We constrain the 2$\sigma$ upper limit to be $\log(\mathrm{H_2O}) \sim -0.5$. This lack of constraint is likely due to the featureless WFC3 spectrum and Spitzer points which lie along the $\sim$2900~K isotherm (see Figure~\ref{fig:wasp18_flux}). This dataset constrains a much more isothermal temperature profile as discussed below in section \ref{sec:wasp18_pt}. H$_2$O thus does not have features in the spectrum regardless of its abundance due to the lack of a significant temperature gradient. Hence the inclusion of H- and dissociation into the model does not help constrain the abundance. In fact, as with the \citet{sheppard2017} dataset, the inclusion of H- and dissociation in these retrievals further increases the uncertainty on the H$_2$O as well as being statistically disfavoured by the Bayesian analysis.

The \citet{arcangeli2019} dataset shows better constraints on the H$_2$O than the \citet{arcangeli2018} data. All of the conducted retrievals showed a peak near the expected solar value, but with a long tail in probability at both low and high abundance meaning that as with the other datasets there is no definitive detection of H$_2$O. The HST WFC3 data does indicate a slightly lower planet/star flux ratio than the other two datasets, with a black body fit closer to 2800~K (Figure~\ref{fig:wasp18_flux}) but this is consistent with the fact that the \citet{arcangeli2019} dataset samples cooler regions of the planetary disk, as stated in section~\ref{sec:intro}. We also see a median P-T profile with an inversion for this dataset but the constraints may also allow for an isothermal/non-inverted profile given the uncertainties (see section \ref{sec:wasp18_pt}). There are a handful of data points at $\sim$1.5~$\mu$m which indicate a potential emission feature from H$_2$O. This small rise in the spectrum near the H$_2$O feature is therefore able to constrain the H$_2$O to a greater degree than for the other two datasets.

\subsection{H- Abundances}\label{sec:h-}

H- also has a non-negligible cross section in the WFC3 bandpass, but we see no conclusive detection in any dataset. Figure~\ref{fig:wasp18histogram} shows the H- abundance for each of the three datasets for the fiducial and wide H- prior cases. Despite the strong cross section of H- in the WFC3 range and a spectral feature at $\sim$1.6~$\mu$m, the H- abundance cannot be well constrained for almost any of the datasets due to the relatively featureless spectra. As thermal dissociation is included the constraint on the H- abundance becomes weaker due to its lower abundance in the upper atmosphere.

\begin{figure*}[ht]
\centering
\includegraphics[width=0.95\textwidth]{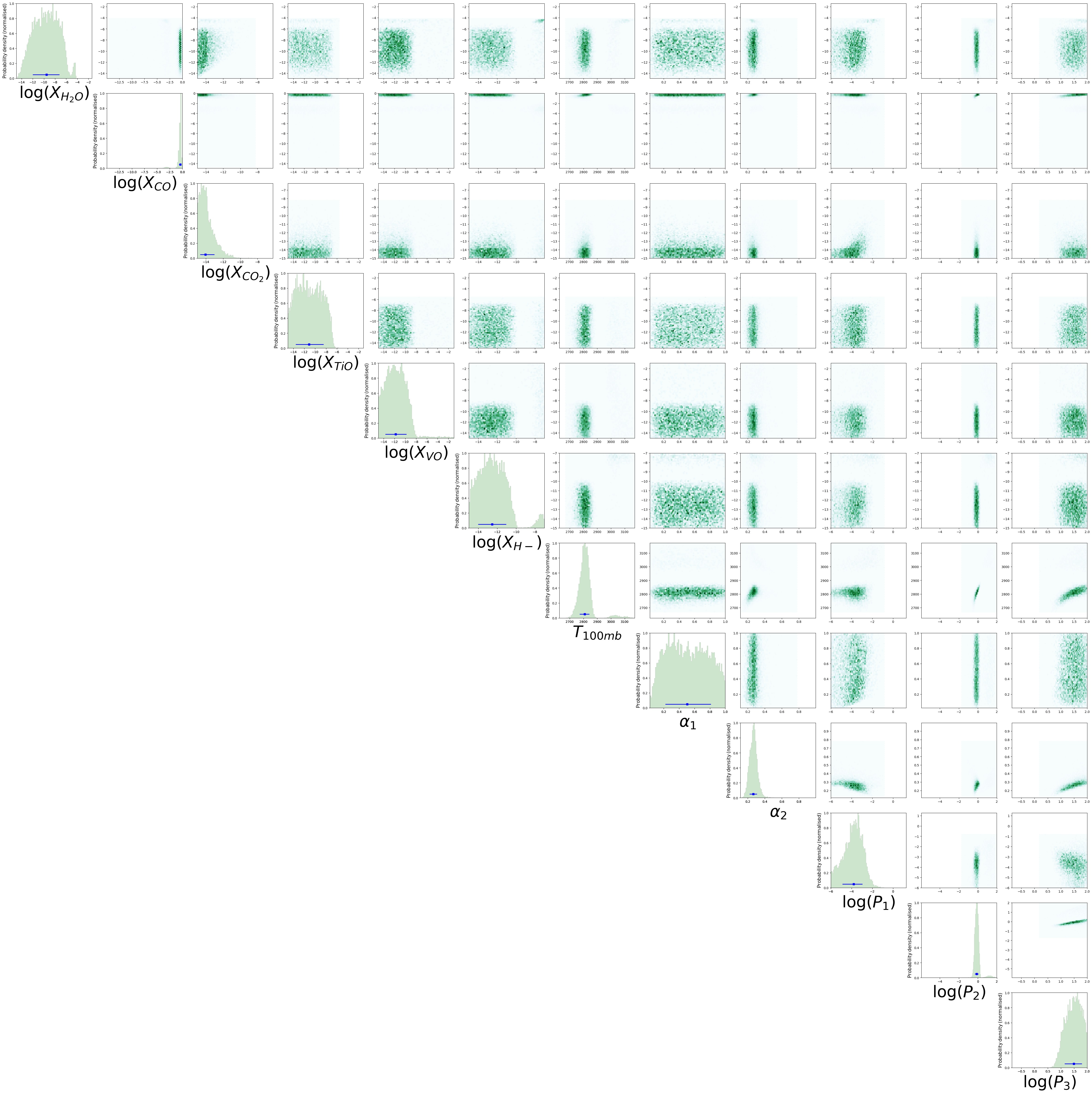}
    \caption{Posterior distribution for the fiducial retrieval of WASP-18b's dayside spectrum from the \citet{sheppard2017} dataset. We retrieve six parameters representing the undissociated abundance of the chemical species and six parameters parametrising the P-T profile as discussed in \citet{madhu2009}.}
    \label{fig:s2017_posterior}
\end{figure*}

Our fiducial model with the \citet{sheppard2017} data shows two distinct modes in the posterior distribution. One mode consists of a super-solar H-, solar CO and sub-solar H$_2$O distribution, while the other suggests a sub-solar H-, super-solar CO and slightly sub-solar H$_2$O. This can be seen in the double peaked posterior distribution of each species in the top panels of Figure~\ref{fig:wasp18histogram}, particularly evident for H-. This arises due to the slight dip in the 1.6~$\mu$m data for the \citet{sheppard2017} dataset shown in Figure~\ref{fig:wasp18_flux}. This feature may be explained by the presence of CO or H-, given that both species have overlapping weak spectral features at this wavelength (see Figure~\ref{fig:wasp18cs}). However, due to the weaker cross section of CO at $\sim$1.6~$\mu$m a very high abundance of $\log(\mathrm{CO})\sim-0.4$ is required for CO to explain the dip in the observations. We show the full posterior distribution for the fiducial retrieval of the \citet{sheppard2017} dataset in Figure~\ref{fig:s2017_posterior}.

\begin{figure*}
	\centering
	\includegraphics[width=\textwidth,trim={0cm 0.0cm 0cm 0cm},clip]{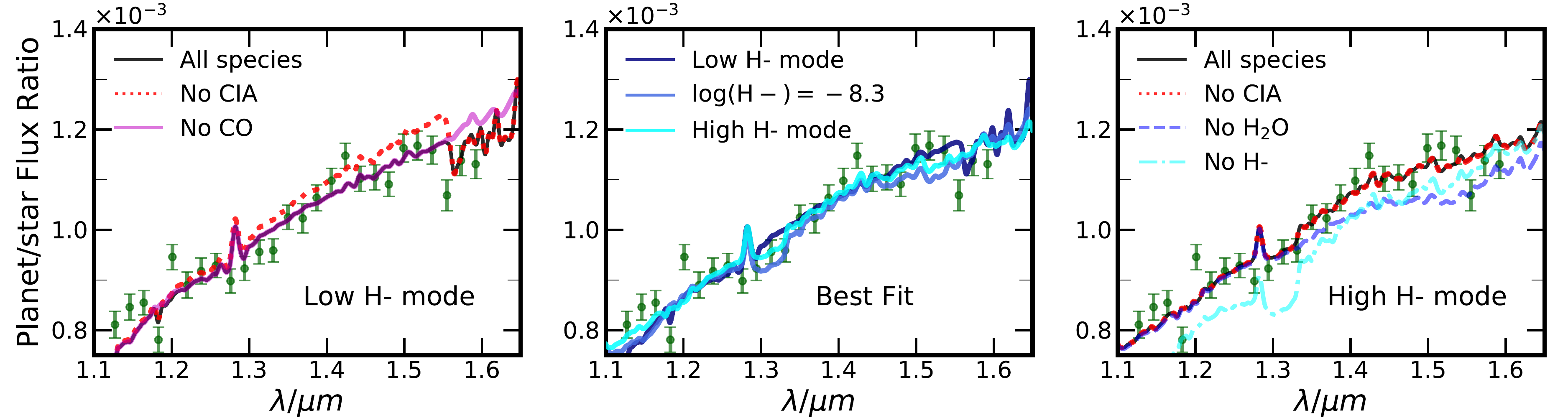}
    \caption{Best fit models for the retrievals of the \citet{sheppard2017} dataset with the different opacity sources removed. The left panel shows the low H- abundance mode and the right panel shows the high H- mode. The middle panel shows the best fit cases for both alongside a best fit model with the H- abundance fixed to the solar equilibrium value \citep{parmentier2018}.}
    \label{fig:best_fit}
\end{figure*}

\begin{figure*}[ht]
\centering
\includegraphics[width=0.95\textwidth]{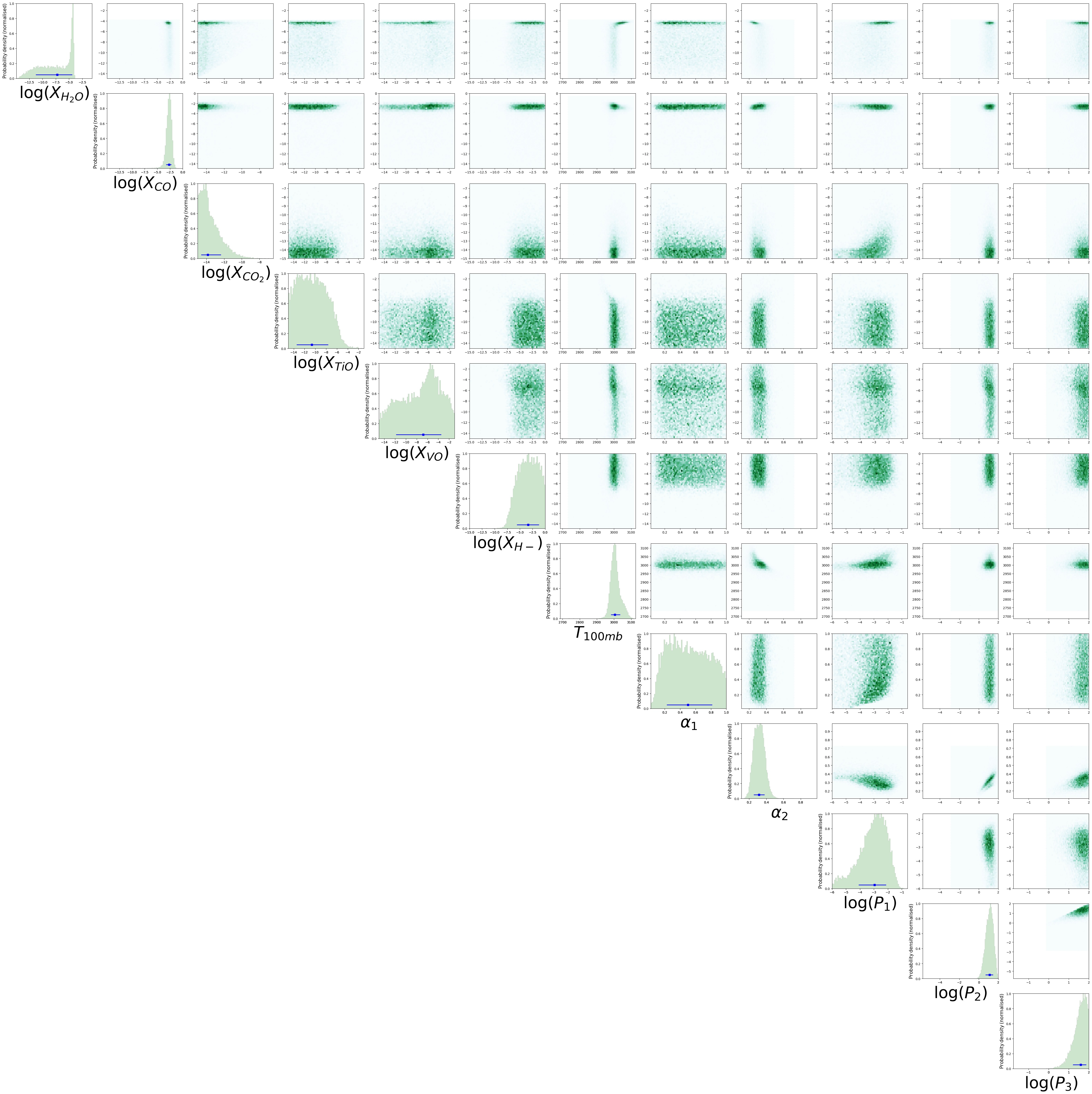}
    \caption{Posterior distribution for the wide H- prior retrieval of WASP-18b's dayside spectrum from the \citet{sheppard2017} dataset. We retrieve six parameters representing the undissociated abundance of the chemical species and six parameters parametrising the P-T profile as discussed in \citet{madhu2009}.}
    \label{fig:s2017_wide_prior}
\end{figure*}

When the H- abundance is unrestricted and allowed to extend up to $\log(\mathrm{H-}) = 0$ for the case of the wide prior, the super-solar H- mode is preferred as shown in Figure~\ref{fig:wasp18histogram} and the posterior distribution in Figure~\ref{fig:s2017_wide_prior}. This is because the H- possesses strong bound-free opacity to explain the feature at the red end of the data. The predicted solar value from \citet{parmentier2018} is $\log(\mathrm{H-}) \approx -8.3$, while our undissociated H- abundance constraint is $\log(\mathrm{H-}) = {-3.4}^{+2.1}_{-2.2}$ as shown in Table \ref{tab:retrievals}. Such an extremely high H- abundance is highly unphysical from chemical models \citep[e.g.][]{parmentier2018}. Therefore the derived H- abundance in the wide prior case is an artefact of the degeneracy with CO given the dependence on few data points near 1.6~$\mu$m. We tested this conclusion by performing a retrieval without the final 4 data points at the red end of the WFC3 and found no constraint on the H- abundance. This super-solar H- mode in the wide prior retrieval is weakly disfavoured over the fiducial case and strongly disfavoured over a model without either H- or dissociation (see Table~\ref{tab:retrievals}). This is because the extra H- parameter introduced into the retrievals does not change the evidence significantly to explain the observations better. In either case the H$_2$O abundance remains sub-solar given that there is no prominent spectral signature in the observations at $\sim$1.4~$\mu$m indicating its presence.

We further investigated the two modes in the posterior distribution for the \citet{sheppard2017} dataset by plotting the best fitting models with varying H- abundances as shown in Figure~\ref{fig:best_fit}. This shows that for the low H- mode, the spectrum is largely determined by the H$_2$-H$_2$/H$_2$-He collisionally induced absorption and the CO absorption. The H$_2$ and CO are present in roughly equal parts at high abundance and thus are the only two species to significantly affect the spectrum. On the other hand the spectrum for the high H- mode is determined by the strong opacity of H- at such high abundances and also partly by the H$_2$O opacity due to the weak peak near $\sim$1.55~$\mu$m. Intermediate H- abundances however are disfavoured as they are unable to match the observations. We note that the narrow peak near $\sim$1.28~$\mu$m is a stellar feature.

\begin{figure*}[ht]
	\includegraphics[width=\textwidth,trim=6.3cm 0 4.6cm 0,clip]{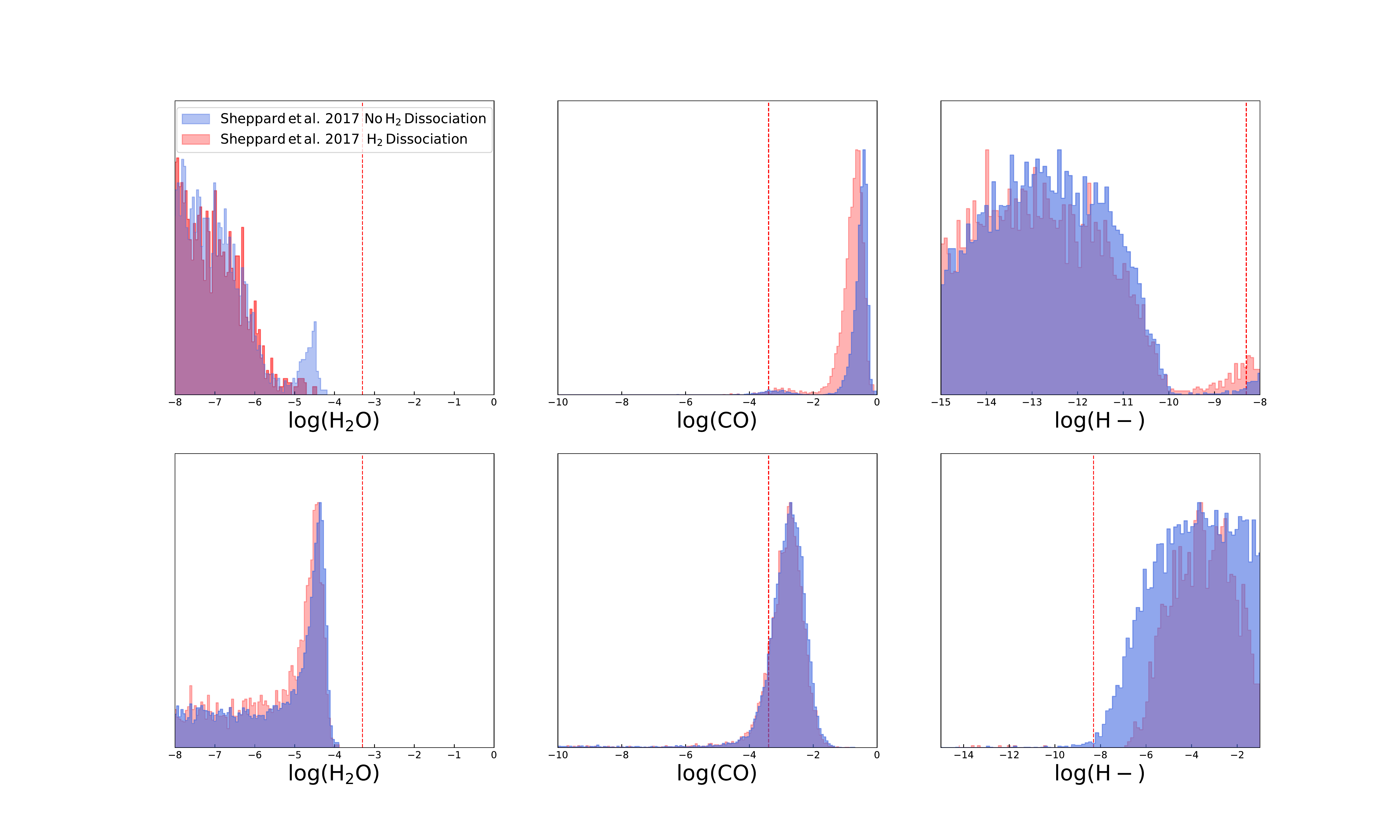}
    \caption{Posterior probability distributions of undissociated H$_2$O, CO and H- abundances for the retrievals of the \citet{sheppard2017} dataset. The red and blue histograms show the retrievals with and without H$_2$ dissociation respectively. The top panels show the fiducial case where the H- prior was restricted to be between $\log(\mathrm{H-}) = -15$ and $-7$, whereas the bottom panel shows the wide prior case where the upper H- prior was extended to $\log(\mathrm{H-}) = -0$. The red dashed line shows the expected abundance for each of the species assuming chemical equilibrium with solar elemental abundances \citep{parmentier2018}.}
    \label{fig:wasp18histogram_h2_dissoc}
\end{figure*}

We have also tested retrievals for the \citet{sheppard2017} dataset which include the dissociation of H$_2$ but we see no significant differences in the retrieved parameters as shown in Figure~\ref{fig:wasp18histogram_h2_dissoc}. This is because the deep atmosphere where the continuum is set is not strongly affected by the dissociation of H$_2$, which occurs more significantly at high altitudes. Therefore the CIA opacity is not significantly altered and thus the spectrum and the bimodal retrieved constraints are largely unchanged. The dissociation of H$_2$ was statistically favoured in the fiducial retrieval but disfavoured in the wide prior retrieval. Thus we cannot conclusively confirm the dissociation of H$_2$ within the atmosphere based on current data.

The other two datasets do not show any constraint for H- for any of the retrievals conducted. These datasets have more isothermal P-T profiles without a potential spectral signature at $\sim$1.6~$\mu$m which makes constraints on H- very weak. We therefore see no strong evidence for H- in any of the datasets, at odds with previous work \citep{arcangeli2018}, which indicated a detection of H- and a thermal inversion from equilibrium models. Our retrievals explored a wide range of temperatures and abundances and found that the featureless HST WFC3 spectrum cannot fully constrain H- definitively. A weak preference for H- without thermal dissociation at 1.1$\sigma$ significance is seen in the \citet{arcangeli2019} dataset (see Table~\ref{tab:retrievals}), but further observations of WASP-18b may be able to provide more stringent constraints on H-.

\subsection{CO Abundances}\label{sec:co}

\begin{figure*}[ht]
    \centering
	\includegraphics[width=0.9\textwidth,trim=2.0cm 0 3cm 0,clip]{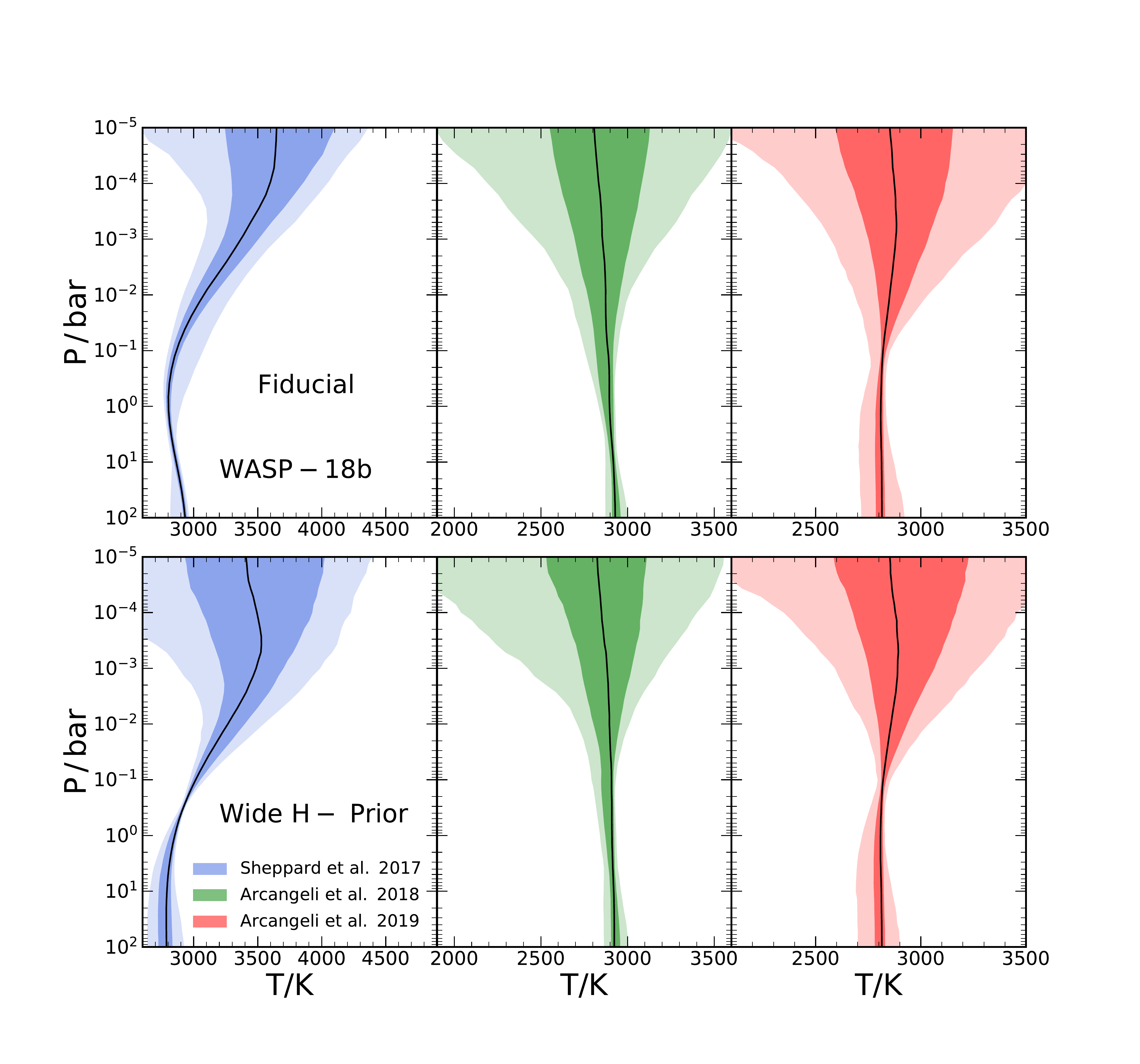}
    \caption{Retrieved pressure-temperature profiles for the retrievals conducted for WASP-18b. The left, middle and right panels show the retrievals for the \citet{sheppard2017}, \citet{arcangeli2018} and \citet{arcangeli2019} datasets respectively. The top panel for each dataset shows the fiducial case and the bottom panels show the case with a wide H- prior. The dark and light colours indicate the 1$\sigma$ and 2$\sigma$ uncertainty respectively, with the solid black line showing the median best fit.}
    \label{fig:wasp18_PT}
\end{figure*}

The Spitzer photometric observations show an emission feature in the 4.5~$\mu$m band for the \citet{sheppard2017} dataset (see Figure~\ref{fig:wasp18_flux}). This is likely due to CO, which has a strong cross section at $\sim4.5\mu$m (see Figure~\ref{fig:wasp18cs}). The retrieved CO abundance is shown in Figure~\ref{fig:wasp18histogram}. The H- and H$_2$O opacities are weaker than that of CO at these wavelengths (see Figure~\ref{fig:wasp18cs}), and thus the 4.5~$\mu$m Spitzer point can only be well explained by the presence of CO. We find a median CO abundance of $\log(\mathrm{CO}) = -0.41^{+0.15}_{-0.28}$ for our fiducial retrieval of the \citet{sheppard2017} dataset including H- and dissociation. The slightly stronger emission feature at 4.5~$\mu$m in the re-analysed Sheppard et al. dataset also constrains a stronger thermal inversion in the photosphere (see Figure~\ref{fig:wasp18_PT}). These results agree well with previous retrievals in \citet{sheppard2017} with one exception. Our wide-prior H- retrieval results in a lower CO abundance of $\log(\mathrm{CO}) = -2.70^{+0.45}_{-0.57}$. This is because the H- abundance provides enough continuum opacity in the 4.5~$\mu$m band to allow for a lower CO abundance to explain the observed flux excess in that band.

The constraints on the CO abundances are weaker for the remaining two datasets due to the larger uncertainties in their 3.6~$\mu$m and 4.5~$\mu$m Spitzer data obtained from \citet{maxted2013}. Our fiducial case constrains CO at $\log(\mathrm{CO}) = -4.0^{+3.6}_{-5.6}$ and $\log(\mathrm{CO}) \lesssim -0.3$ for the \citet{arcangeli2018} and \citet{arcangeli2019} datasets respectively. The retrievals on the \citet{arcangeli2018} dataset do still indicate a modal CO value at super-solar abundance ($\log(\mathrm{CO})\sim-1$) (see Figure~\ref{fig:wasp18histogram}), but the uncertainty is much greater and the distributions have long tails extending to significantly lower abundances. Therefore, the median abundances and uncertainties retrieved allow for solar as well as super-solar CO abundances, as shown in Table \ref{tab:retrievals}. The retrieved CO abundances are largely unaffected by the inclusion of thermal dissociation and H- opacity. This is unsurprising given that CO does not thermally dissociate at such temperatures and any H- present would only contribute relatively weaker free-free absorption in the Spitzer bandpass at physically plausible quantities (see Figure~\ref{fig:wasp18cs}). The \citet{arcangeli2019} dataset does not show any significant constraints on the CO as the Spitzer observations from \citet{maxted2013} largely follow a black body spectrum (see Figure~\ref{fig:wasp18_flux}).

We tested whether the Spitzer data was indeed the source of the CO abundance and thermal inversion constraints by retrieving each dataset with only the HST WFC3 observations. All except one showed a temperature profile that was isothermal, given that the WFC3 spectrum is largely featureless, and no constraint on any species. The \citet{sheppard2017} dataset was the outlier and showed a bi-modal CO constraint due to the slight downturn in the WFC3 points near 1.6~$\mu$m, similar to that seen in the fiducial case discussed previously in section \ref{sec:h-}. The CO has molecular absorption in the WFC3 range near the H- feature at $\sim$1.6~$\mu$m (see Figure~\ref{fig:best_fit}) and therefore it is not unexpected that the dip in the WFC3 points may be explained by either species. Our retrievals show evidence for either CO or H-, the former of which was claimed in \citet{sheppard2017} given that H- was not included in their analysis. This constraint on CO is however much weaker than with the inclusion of the Spitzer photometric bands, where CO has a much stronger feature near $\sim$4.5~$\mu$m (see Figure~\ref{fig:wasp18cs}). The arrival of JWST is therefore likely to place much stronger constraints on the CO abundance due to its higher precision and higher-resolution observations covering this CO absorption feature.

\subsection{P-T Profile}\label{sec:wasp18_pt}

Figure~\ref{fig:wasp18_PT} shows the P-T profiles for the fiducial and the wide H- prior retrievals for each dataset. Our work shows the best constraints on the P-T profile occur in the photosphere probed at P$\sim$0.1~bar, where most of the emission occurs. Lower pressures (P$\lesssim10^{-3}$~bar) show a much less constrained temperature given that they have a negligible effect on the spectrum. The effect is similar at higher pressures (P$\gtrsim$10~bar) where the onset of the isotherm and the high optical depth limit our ability to constrain the P-T profile. However, there are some key differences in the retrieved P-T profiles which we explore below.

The \citet{sheppard2017} retrievals show a thermal inversion present in the photosphere (P$\sim$100~mbar) with a T$_\mathrm{100mb}\sim 2800-3000$~K. This inversion is primarily constrained by the Spitzer photometric data which show a strong CO emission feature at 4.5~$\mu$m. We see an inversion present in every retrieval run for this dataset, with and without H- or dissociation. This dataset shows the strongest thermal inversion of all three datasets with a temperature increase of $\sim$600~K going upwards in the atmosphere as shown in Figure~\ref{fig:wasp18_PT}. However, our results for \citet{arcangeli2018} indicate a more isothermal P-T profile. This is caused by the weaker Spitzer constraints from the \citet{maxted2013} observations with larger error bars as well as the featureless WFC3 data. These points all mostly lie along the isotherm of $\sim$2900~K and hence we do not see much variation whether H- or dissociation is included. Equilibrium models from \citet{arcangeli2018} indicated the presence of a thermal inversion in the upper atmosphere, and our results do allow for a thermal inversion at pressures $\lesssim$10$^{-2}$~bar given that the temperature is less well constrained in the upper atmosphere. Hence despite the largely isothermal profile, the large uncertainty does mean that we constrain the P-T profile to be within 2$\sigma$ of that reported in \citet{arcangeli2018}.

The spectra of \citet{arcangeli2019} showed an inversion similar to but weaker than \citet{sheppard2017} due to the less well constrained emission features from CO in the Spitzer 4.5~$\mu$m band. The slightly lower planet/star flux ratio in the WFC3 and Spitzer bands also constrains a slightly cooler temperature at T$_\mathrm{100mb}\sim$2800~K as shown in Table~\ref{tab:retrievals}, which may be ascribed to the different phase sampling for this dataset as discussed in section~\ref{sec:intro}. Generally, we see a 100~mbar temperature that is very close to the expected dayside photospheric temperature of WASP-18b without significant redistribution for all three of the datasets.

We additionally tested whether our assumed stellar temperature (given in Table \ref{tab:wasp18_params}) affected the results by varying the stellar temperature by 100~K (the 1$\sigma$ error in its value) and rerunning our retrievals. We found no significant change to the retrieved results for any of the parameters. We did see a change in the peak in the distribution for the 100~mbar temperature by $\sim$45~K, as expected in order to match the observed planet/star flux ratio in the dataset.

\subsection{Constraints on other species}

\begin{figure*}[ht]
\centering
\includegraphics[width=0.95\textwidth]{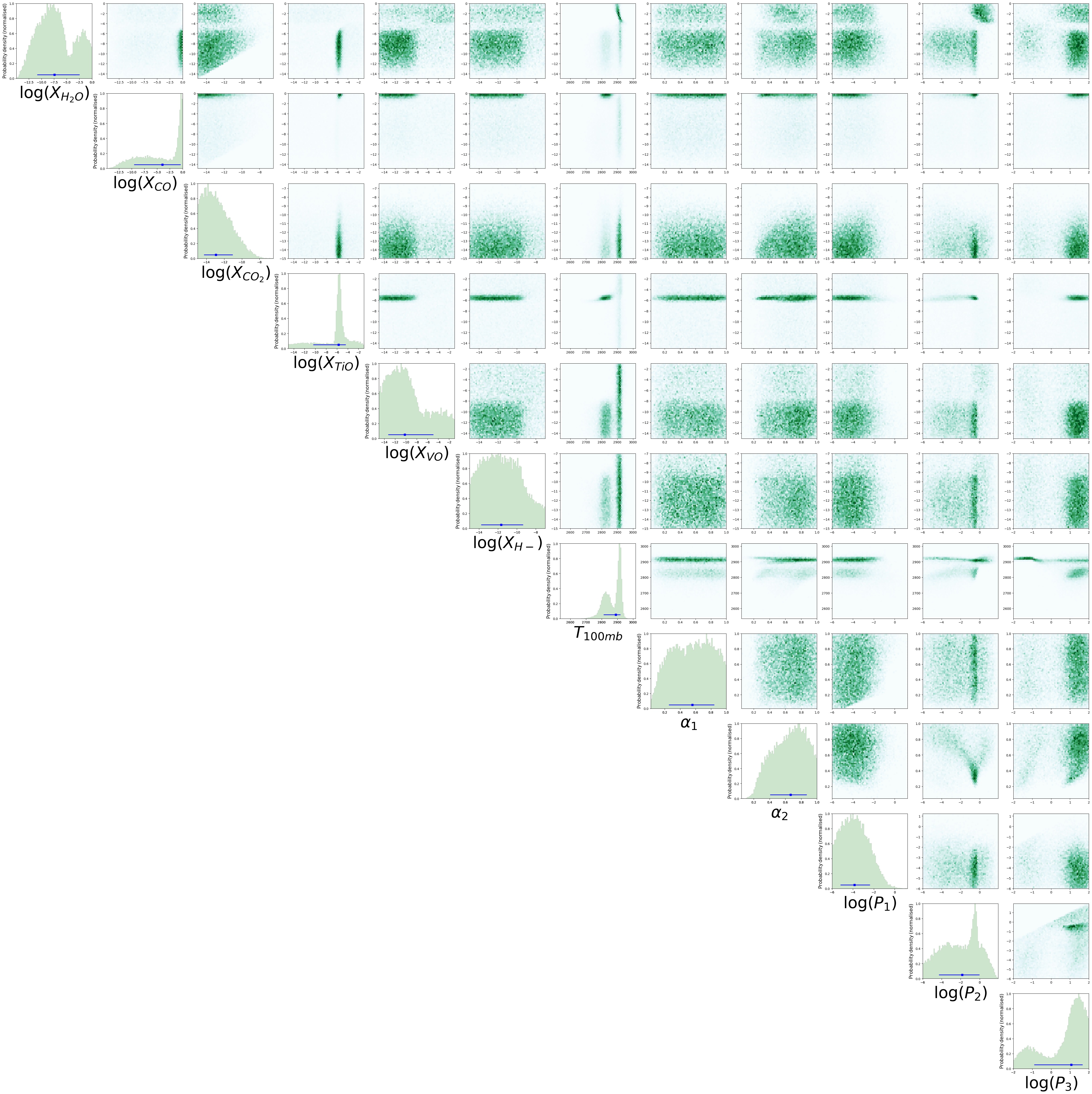}
    \caption{Posterior distribution for the fiducial retrieval of WASP-18b's dayside spectrum from the \citet{arcangeli2018} dataset. We retrieve six parameters representing the undissociated abundance of the chemical species and six parameters parametrising the P-T profile as discussed in \citet{madhu2009}.}
    \label{fig:a2018_posterior}
\end{figure*}

\begin{figure*}[ht]
\centering
\includegraphics[width=0.95\textwidth]{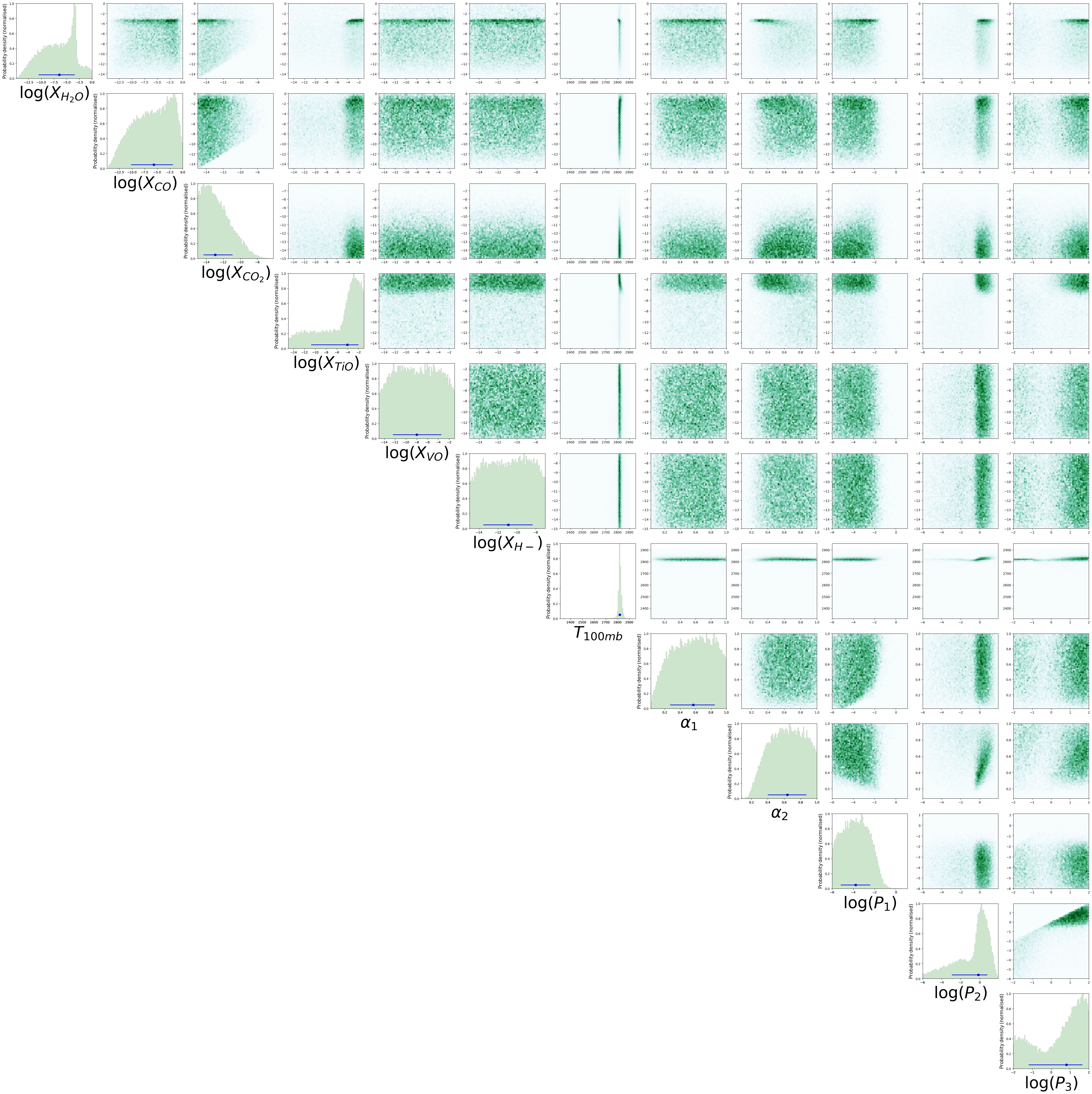}
    \caption{Posterior distribution for the fiducial retrieval of WASP-18b's dayside spectrum from the \citet{arcangeli2019} dataset. We retrieve six parameters representing the undissociated abundance of the chemical species and six parameters parametrising the P-T profile as discussed in \citet{madhu2009}.}
    \label{fig:a2019_posterior}
\end{figure*}

Our retrievals are unable to constrain TiO or VO significantly. Thermal inversions have been predicted from equilibrium models of hot Jupiters due to species such as TiO and VO \citep{fortney2008, spiegel2009}. In our retrievals we only observe very weak constraints for TiO using the \citet{arcangeli2018} dataset, which shows a peak at $\log(\mathrm{TiO}) \sim -6$ (see Figure~\ref{fig:a2018_posterior}) but with a long-tailed distribution at low and high abundance. The \citet{arcangeli2019} dataset also showed a broad peak at unphysically high abundances of $\log(\mathrm{TiO}) \sim -3$, as shown in the posterior distribution in Figure~\ref{fig:a2019_posterior}. This also has a large uncertainty and tail to lower values. Expectations from solar abundances \citep{asplund2009} and equilibrium models of WASP-18b \citep{parmentier2018, lothringer2018} indicate that these species should be present at $\log(\mathrm{TiO}) \sim -7$ and $\log(\mathrm{VO}) \sim -9$ \citep{asplund2009}. The atmosphere of WASP-18b may also be more susceptible to thermal inversions from species such as TiO and VO if the H$_2$O abundance is sub-solar as indicated by the \citet{sheppard2017} data or thermally dissociated. This is because the presence of strong infrared opacity (such as H$_2$O) can act to cool the upper atmosphere due to its radiative efficiency, thereby reducing the heating effectiveness of species such as TiO and VO \citep{molliere2015, gandhi2019}. Therefore a low H$_2$O abundance makes thermal inversions more likely.

We also do not constrain the CO$_2$ abundance significantly. Previous theoretical studies have also shown CO$_2$ is rarely dominant over H$_2$O and CO in hot Jupiter atmospheres \citep{madhu2012, moses2013, heng2016}. We thus limit our model to include $\log(\mathrm{CO_2}) \leq \log(\mathrm{CO})$ and $\log(\mathrm{CO_2}) \leq \log(\mathrm{H_2O})$. This also allows us to break the degeneracy between CO and CO$_2$ in the 4.5~$\mu$m Spitzer band, where both of these species have strong absorption \citep{madhu2010, kreidberg2014, gandhi2018}.

\section{Summary and Discussion}\label{sec:wasp18_conclusion}

We present here a case study of retrieval analysis for ultra-hot Jupiter (UHJ) atmospheres with equilibrium temperatures in excess of 2000~K. We focus on the UHJ WASP-18b. The motivation for this work arises from recent work on equilibrium models of such ultra-hot planets which show that H- opacity and thermal dissociation of species such as H$_2$O may be important \citep{parmentier2018, lothringer2018, arcangeli2018, kreidberg2018, mansfield2018}. To achieve this we have extended our recently developed HyDRA model \citep{gandhi2018} in order to retrieve the dissociated abundances of numerous species as well as the H- mixing ratio within the atmosphere. To study the effect of H- and dissociation we carry out five separate retrievals on each of the three available datasets with dayside observations of the hot Jupiter WASP-18b \citep{sheppard2017, arcangeli2018, arcangeli2019}.

We do not find strong evidence for H- in any of the datasets. The retrievals of the \citet{arcangeli2018} and \citet{arcangeli2019} datasets show no significant constraints on any species when H- and dissociation are included (see Figure~\ref{fig:wasp18histogram}). On the other hand, the \citet{sheppard2017} dataset does show two distinct families of solutions depending on the H- prior. With a wide H- prior the data favour a super-solar H- abundance of $\log(\mathrm{H-}) = -3.4^{+2.1}_{-2.2}$, which is $\sim$4-6 orders of magnitude greater than predicted from dayside equilibrium models of WASP-18b \citep{arcangeli2018}. However, our fiducial case with an H- prior restricted to more realistic values ($\log(\mathrm{H-}) \lesssim -7$) constrains only an upper limit on the H- that is consistent with solar abundance. However, the fiducial retrieval constrains a super-solar CO abundance of $\log(\mathrm{CO}) = -0.41^{+0.15}_{-0.28}$. This value is between $\sim$2-4 orders of magnitude greater than solar abundance \citep{madhu2012, moses2013}, which also may not be realistic.

We are therefore currently left with two possibly unphysical families of solutions, and must wait for improved data in the future to narrow our solution space. However, for almost all retrievals for the currently available datasets, the inclusion of H- and thermal dissociation is statistically disfavoured as shown in Table~\ref{tab:retrievals}. We thus do not see any strong evidence for H- in any of the dayside observations for WASP-18b, counter to that reported in \citet{arcangeli2018}. However, we do suggest that H- be included as a retrieval parameter for analyses of UHJs (T$_\mathrm{eq} \gtrsim$2000~K) despite the fact that we are unable to detect it given that it does have the potential to influence the spectrum in the HST WFC3 range \citep[e.g.][]{arcangeli2018, parmentier2018, lothringer2018, mikal-evans2019}.

Without any significant spectral features in the WFC3 observations, we find that H$_2$O is also undetected in our current work. We retrieved the H$_2$O abundances for each of the datasets, with and without H- or dissociation. We saw good agreement between these, as shown in Table \ref{tab:retrievals}. We constrain an upper limit for H$_2$O that is slightly sub-solar for the \citet{sheppard2017} dataset, even with the inclusion of dissociation and H- opacity. The \citet{arcangeli2019} dataset similarly indicates a weak peak for H$_2$O, but closer to solar composition. The \citet{arcangeli2018} data on the other hand do not show significant H$_2$O constraints given the more isothermal temperature profile that is retrieved (see Figure~\ref{fig:wasp18_PT}). We therefore conclude that the inclusion of H- and/or thermal dissociation in our retrievals does not help to fully constrain the H$_2$O abundance. Further observations, either with improved wavelength coverage and sensitivity or perhaps through transmission spectroscopy, may be required to resolve this conundrum.

We find different constraints on the CO abundance depending on the datasets. Retrievals with the \citet{sheppard2017} dataset were the only ones which showed a clear detection peak for CO due to the smaller error bars on the Spitzer data. The expected CO abundance from chemical equilibrium models is $\log(\mathrm{CO}) \approx -3.4$ \citep{madhu2012, moses2013}, and our retrievals constrained CO to be between $\log(\mathrm{CO}) = -2.70^{+0.45}_{-0.57}$ and $\log(\mathrm{CO}) = -0.41^{+0.15}_{-0.28}$ depending on the H- prior (see Table~\ref{tab:retrievals}). This is due to its degeneracy with H- as discussed in section \ref{sec:h-}. A super-solar CO abundance was reported in \citet{sheppard2017} from retrievals without H-, and \citet{arcangeli2018} used equilibrium models to argue the importance of H- opacity and obtained a solar CO abundance. Our retrievals indicate that both proposed explanations are able to fit the observations for the \citet{sheppard2017} data, but one of either CO or H- has to be super-solar. Neither of the other two datasets showed any significant constraints on CO, with only a weak super-solar peak seen in the retrievals of the \citet{arcangeli2018} dataset.

Determining the CO abundance is crucial as this has been shown to be a good proxy for the metallicity \citep{madhu2012}. It does not vary significantly with the atmospheric C/O ratio for temperatures in excess of $\sim$1500~K and does not thermally dissociate at the temperatures typical for UHJs \citep[e.g.][]{moses2013, parmentier2018}. For a solar composition atmosphere the CO abundance is $\log(\mathrm{CO}) \approx -3.4$ \citep{madhu2012, moses2013}. The super-solar CO abundances seen for some of the retrievals for the \citet{sheppard2017} dataset may thus indicate a very high metallicity, while others are consistent with near-solar abundance. Additionally, the retrievals with the \citet{arcangeli2018} and \citet{arcangeli2019} datasets are also consistent with solar metallicity given their larger uncertainties. Hence, future work that can improve the CO abundance constraints may be able to better constrain the metallicity of WASP-18b.

The stronger 4.5~$\mu$m emission feature in the Spitzer observations also constrained a thermal inversion in the \citet{sheppard2017} dataset as shown in Figure~\ref{fig:wasp18_PT}. The \citet{arcangeli2018} on the other hand constrained a much more isothermal temperature profile given that the Spitzer data had larger error bars and the HST WFC3 spectra are all largely featureless and lie along the same isotherm (T$\sim$2900~K). The \citet{arcangeli2019} dataset showed evidence for a thermal inversion but with weaker constraints than the \citet{sheppard2017} dataset.

We note that our retrievals do differ from the equilibrium models used in \citet{arcangeli2018} for their inference of H-. Our work assumes that the abundances of the chemical species can be treated as free parameters whereas the equilibrium models used in \citet{arcangeli2018} assume that the abundances are in chemical equilibrium and can be derived from the stellar redistribution, metallicity and C/O ratio. Hence our retrievals allow for chemical species not in thermochemical equilibrium. In addition, \citet{arcangeli2018} use radiative-convective equilibrium P-T profiles which indicate the presence of thermal inversions in the upper atmosphere. We parametrise the P-T profile which allows us to explore a wide range of inverted, non-inverted and isothermal profiles. Our retrievals indicate more isothermal or weakly inverted temperature profiles due to largely featureless data (see Figure~\ref{fig:wasp18_PT}). A precise balance of the overall infrared to visible opacity is required to achieve isotherms in an atmosphere \citep{guillot2010, molliere2015, gandhi2019} which may be unlikely for UHJs \citep[e.g.][]{lothringer2018}. In reality we would expect both approaches of retrievals and grids of equilibrium models to converge to the same results in the presence of well constraining data with strong spectral features, as was achieved on observations of WASP-43b \citep{gandhi2018}. 

We should also highlight the assumptions within the dissociation model used in the retrievals. Firstly, we have assumed that the ionisation of H- may be parametrised in a similar way to the other species. In reality, the H- abundance is set by complex chemical interactions and strongly dependent on the free electron abundance \citep{parmentier2018}. Hence this assumption breaks down away from solar composition. We use this parametric model for H- so we can perform a free chemical retrieval due to lack of a better prescription in such a case. A potential solution is to perform retrievals in chemical equilibrium and vary the C, O and N abundance as was recently done for WASP-121b \citep{mikal-evans2019}.

We also leave the abundance of each species as a free parameter in our retrieval and assume that only this abundance and the temperature profile affect the thermal dissociation of each species. The dissociation reactions only involve the corresponding species and their dissociated byproducts \citep{parmentier2018}, and are unaffected by the abundances of other species. Thus, we can justify assuming the dissociating species are in chemical equilibrium with respect to thermal dissociation, even though we do not assume them to be in chemical equilibrium with respect to other species in the atmosphere.

This work represents a crucial step in the characterisation of UHJs, which have recently come to the fore thanks to high precision observations \citep[e.g.][]{haynes2015, evans2017, sheppard2017, kreidberg2018}. With such extreme temperatures our understanding of atmospheric processes has been pushed to the limit. For instance, theoretical models of planets with extreme irradiation with equilibrium temperatures well in excess of 2000~K are now being explored \citep[e.g][]{parmentier2018, kitzmann2018}. The high temperature HyDRA retrieval framework may form the basis for compositional studies of UHJs as more observational data becomes available. WASP-18b is a JWST early release science target \citep{bean2018} and thus we may soon be able to shed more light on these physical processes for UHJs and finally confirm the presence of H- and/or dissociation in the dayside atmosphere.

\acknowledgments
SG acknowledges support from the UK Science and Technology Facilities Council (STFC). We thank the anonymous referee for a careful review of our manuscipt.

\vspace{5mm}

\bibliography{references}
\bibliographystyle{yahapj}

\end{document}